\documentstyle[12pt,preprint,eqsecnum,aps,prd]{revtex}

\input epsf
\begin{document}

\preprint{\vbox{\hbox{MADPH-99-1098} 
                \hbox{FERMILAB-PUB-99/028-T}}}
 
\title{Curved QCD string dynamics} 
\author{Theodore J. Allen}
\address{Department of Physics, Eaton Hall \\
Hobart and William Smith Colleges \\
Geneva, New York 14456 USA }

\author{M. G. Olsson}
\address{Department of Physics, University of Wisconsin, \\
1150 University Avenue, Madison, Wisconsin 53706 USA }

\author{Sini\v{s}a Veseli}
\address{Theoretical Physics Department \\
Fermi National Accelerator Laboratory \\
P.O. Box 500, Batavia, Illinois 60510 USA}

\date{\today}

\maketitle

\thispagestyle{empty}
\vskip -20 pt

\begin{abstract}
We consider the effects of going beyond the approximation of a
straight string in mesons by using a flexible flux tube model wherein a
Nambu-Goto string bends in response to quark accelerations.  The curved
string is dynamically identical to the straight string even for
ultra-relativistic mesons except for a small additional radial momentum.
We numerically solve the curved string model in the case where
both ends have equal mass quarks and also the case where one end is fixed.
No approximation of non-relativistic motion is made.  We note some small
but interesting difference from the straight string.
\end{abstract}

\newpage

\section{Introduction}\label{sec:intro}

As the separation between quarks becomes large, their color field contracts
into a string-like or flux tube configuration.  QCD is thus thought to
resemble string theory in the large distance regime.  The
Nambu-Goto-Polyakov QCD string \cite{nambugoto} is an elegant model for
this color field from which valuable insights on the nature of QCD are
expected.  The most obvious of these is the well known picture of static
linear confinement.  For a dynamical hadron the simplest assumption for the
long distance color field configuration is that the flux tube/string always
remains straight.  This was the assumption of the relativistic flux tube
model for mesons \cite{aft} which has enjoyed phenomenological success.

As pointed out by Nesterenko \cite{nesterenko} such rigid string models
cannot strictly be considered Nambu-Goto strings since the string cannot
remain straight during dynamical motion of the quarks.  Recently, the
present authors \cite{adiabatic} found an explicit solution to the string
equations describing the deviation from straightness due to non-uniform end
motion.  The principal assumptions in our calculation are to treat the
motion as ``adiabatic'' and that the deviation from straightness is
small. The adiabatic assumption is just that the string shape depends on
the end motion and that there is no other explicit time dependence.  By
direct calculation we find that for all physical cases the string never
bends very much even for relativistic motion.

An interesting conclusion of our previous work \cite{adiabatic} was that
almost every dynamical quantity for the curved string is the same as the
straight string.  The angular momentum, energy, and momentum perpendicular
to the line connecting the quarks are unchanged for small string
deformations assuming the quark distance and angular velocity are the same
for the curved and straight strings.

The sole difference between the dynamically curved string and the straight
string is the component of momentum along the line connecting the quarks.
From Fig.~1 the origin of this radial momentum is clear.  It is also
evident that the contribution from the outer part of the string will
predominate due to the larger velocity of the end of the string. The net
result will be an inward flowing radial momentum for the case shown in
Fig.~1 corresponding to increasing angular velocity ($\dot\omega > 0$).
For a meson the total angular momentum is conserved and hence if the
angular velocity is increasing the distance between the quarks must be
decreasing.  We conclude that the tube radial momentum due to curvature
always has the same sign as the quark's radial momentum.  In a meson with
equal mass quarks the two halves of the string have equal but opposite
radial momenta, just as the quarks do.

Since the curved string meson is nearly identical to the rigid (straight)
string meson similar methods are indicated for our numerical solution.  The
effect of the curvature induced radial momentum is to shift the energy
level solutions downward only a few MeV.  For practical purposes it is
quite sufficient to ignore deviation from straightness even for massless
quarks with resulting ultra-relativistic kinematics.  A principal result
thus is that one can use the simpler straight string approximation with
negligible loss in accuracy.

The remainder of the paper is organized as follows.  In Sec.\
\ref{sec:action} we start with the string plus quark action and establish
the equations of motion and conserved quantities.  The solution for the
curved string is reviewed in Sec.\ \ref{sec:curved} and the energy,
angular momentum, and linear momentum are computed.  We show here that the
curvature of the string induced by general quark motion does not contribute
to the energy and angular momentum.  The sole effect of the curved string
is to enhance the radial momentum of the quark.  The method of quantization
of the quark-string system and the solutions scheme is covered in Sec.\
\ref{sec:numerical}.  Our numerical results and general conclusions are
given in Sections \ref{sec:results} and \ref{sec:conclusion}.

\section{The string-quark action}\label{sec:action}

We begin by reviewing the formalism of the Nambu-Goto-Polyakov
\cite{nambugoto,barbashovnesterenko} string and spinless quark actions and
some expressions for conserved quantities.  The action for a quark and
anti-quark connected by a QCD string is
\begin{equation}
S= -\,{a\over 2}\int d\tau\int d\sigma\, \sqrt{-h}\, h^{ab} X^\mu_{,a}
X^\nu_{,b}\,\, \eta_{\mu\nu}\ - \sum_{i=1,2} \int d\tau\,
\left(m\sqrt{-\dot{x}^\mu\dot{x}_\mu}\right)_i \, , \label{action}
\end{equation}
where $h^{ab}$ is the inverse of the two-dimensional metric $h_{ab}$ whose
indices run over $\tau$ and $\sigma$, while $h=\det(h_{ab})$.
$X^\mu(\tau,\sigma)$ is the string position and $X^\mu_{,a} \equiv
\partial_a X^\mu$.  The two quark coordinates are $x^\mu_i(\tau) \equiv
X^\mu(\tau,\sigma_i)$ at the ends of the string.  We use the metric
signature conventions
that $h = (-, +)$ and $\eta = (-, +, +, +)$.

The action (\ref{action}) is classically equivalent to the string tension
$a$ times the worldsheet area swept out by the string minus the masses of
the quarks times their worldline lengths.  Variation of the Lagrange
multiplier $h_{ab}$ determines that it is equal to the embedding metric, up
to a multiplicative constant
\begin{equation}
h_{ab} = X^\mu_{,a}X^{\nu}_{,b}\,\eta_{\mu\nu}\, . \label{metric}
\end{equation}
Variation of the string and quark positions yields the equations of motion
\begin{eqnarray}
(\sqrt{-h}\,\,h^{ab}\, X^\mu_{,b})_{,a} &=& 0\quad ,  \label{StringEoM} \\
-\, (-)^i a\, \sqrt{-h}\, h^{\sigma b}\, X^\mu_{,b} &=&
(p_i^\mu)_{,\tau}\quad , \label{QuarkEoM}
\end{eqnarray}
where the canonical quark momenta are
\begin{equation}
p_i^\mu = {\delta S_{\rm quark} \over \delta \dot{x}^\mu_i} =
\left({m\dot{x}^\mu\over \sqrt{-\dot{x}^\nu\dot{x}_\nu}}\right)_i\equiv
(m\gamma\dot{x}^\mu)_i\, ,
\end{equation}
and $\dot{x_i}^\mu = (x^\mu_i)_{,\tau}$.  The string momentum density is
\begin{equation}
\Pi_\mu(\tau,\sigma) = {\delta S_{\rm string} \over \delta \dot{X}^\mu} =
-a \sqrt{-h}\, h^{\tau a}\, X_{\mu\, ,a}(\tau,\sigma)\, .
\end{equation}
The total energy and momentum of the string-quark system then follows
\begin{equation}
P^\mu_{\rm total} = P^\mu_{\rm string} + P^\mu_{\rm quarks} =
\int_{\sigma_1}^{\sigma_2} \Pi^\mu (\tau,\sigma)\, d\sigma + \sum_{i=1,2}
p^\mu_i\, ,
\end{equation}
and the system angular momentum is 
\begin{equation}
J^{\mu\nu}_{\rm total} = J^{\mu\nu}_{\rm string} + J^{\mu\nu}_{\rm quarks} =
\int_{\sigma_1}^{\sigma_2} X^{[\mu}\,\Pi^{\nu]}\, d\sigma + \sum_{i=1,2}
\left(x^{[\mu}\, p^{\nu]}\right)_i\, .
\end{equation}

Because of the spherical symmetry of the string and quarks system, we may
consider motion in the $(x^1,x^2)$ plane. It is useful then to use helicity
coordinates
\begin{equation}
X^\pm = {X^1 \pm i X^2 \over\sqrt 2}\, .
\end{equation}
We will adopt the notation
\begin{equation}
{\bf X} \equiv X^+\, , 
\end{equation}
and therefore ${\bf X}^* = X^-$. In terms of the spatial coordinate ${\bf
X}$, the above results become
\begin{eqnarray}
h_{ab} &=& - X^0_{,a}X^0_{,b}+2\,\,{\rm Re}\,({\bf X}^*_{,a}{\bf X}_{,b})\ ,
\label{hab} \\ E &=& P^0 = - a \int_{\sigma_1}^{\sigma_2} d\sigma \,
\sqrt{-h}\, h^{\tau a} X^0_{,a} + \sum_{i=1,2} \left( m\gamma\right)_i\,
.\label{stringE}
\end{eqnarray}
The spatial momentum can be expressed in helicity form as
\begin{eqnarray}
{\bf P} &=& {1\over\sqrt2}\left(P^1 + i P^2\right) \nonumber \\
&=& - a \int_{\sigma_1}^{\sigma_2} d\sigma \, \sqrt{-h}\, h^{\tau a}
{\bf X}_{,a} + \sum_{i=1,2}\left( m\gamma\dot{{\bf x}} \right)_i\, ,\label{spatialP}
\end{eqnarray}
where the quark coordinate ${\bf x}_i(\tau)$ is
\begin{equation}
{\bf x}_i(\tau) = {\bf X}(\tau,\sigma_i)\, .
\end{equation}
Similarly, the angular momentum $J^3$ perpendicular to the plane of motion is
\begin{equation}
J^3 = 2 a\int_{\sigma_1}^{\sigma_2} d\sigma \, \sqrt{-h}\, h^{\tau b}\,
{\rm Im}({\bf X}^*_{,b}\, {\bf X}) + 2 \sum_{i=1,2}\left( m\gamma\, {\rm
Im}(\dot{{\bf x}}^* {\bf x})\right)_i\, .
\end{equation}

\section{The curved string with one fixed end}\label{sec:curved}

As we showed previously, it is possible to perturb about an exact straight
string solution to obtain the shape of a string in which one end is fixed
and the other has a sufficiently small angular acceleration.  The string
motion in a meson can be considered as a special case where the system
angular momentum and energy are conserved. We begin this section by
re-establishing this result by an abbreviated method.  In \cite{adiabatic}
we began with the ansatz
\begin{equation}
\begin{array}{rcl}
X^0(\tau,\sigma) &=& \tau\ =\ t\ , \\ {\bf X}(\tau,\sigma) &=& {1\over \sqrt2}\left(\sigma
R(t) + F(\sigma)\right)\, \exp{\left[i(\omega t + \phi(t))\right]}\ , \\
\end{array}
 \label{perturbX}
\end{equation}
where $F(\sigma)$ is complex and $F(\sigma)$ and $\phi(t)$ are small.
A more convenient starting point can be found by observing that if
$X(t,\sigma=1) \equiv {R\over\sqrt2}$, reparametrization invariance allows
us to set ${\rm Re}(F) = 0$.  We can then exploit the assumption of small
deviation from straightness to make the simpler but related ansatz 
\begin{equation}
\begin{array}{rcl}
X^0(\tau,\sigma) &=& \tau\ =\ t\, , \\
{\bf X}(\tau,\sigma) &=& {1\over\sqrt 2}\sigma R(t)\, \exp{(i\Phi)}\, ,
\end{array}
\label{expX}
\end{equation}
where the overall phase is a rigid rotation plus the sum of two
phases, each dependent upon only one of the coordinates
\begin{eqnarray}
\Phi &=& \omega t + \phi(t) + f(\sigma) \, .
\end{eqnarray}
With $X^\mu$ of the above form, from Eq.~(\ref{hab}) we see that the
worldsheet metric is given as
\begin{eqnarray}
h_{ab} &=& -\delta_{at}\delta_{tb} + (\sigma R)_{,a}(\sigma R)_{,b} +
(\sigma R)^2 \Phi_{,a}\Phi_{,b}\,\, ,
\end{eqnarray}
with individual components
\begin{eqnarray}
h_{tt}&=&-\,\gamma^{-2}(\sigma) + 2\sigma^2 R^2 \omega\dot\phi\, , \\
h_{t\sigma}&=&\sigma R\dot{R} + \sigma^2 R^2\omega f^{\prime}\, , \\
h_{\sigma\sigma}&=& R^2\, ,
\end{eqnarray}
and volume density 
\begin{eqnarray}
\sqrt{-h} &=& \sqrt{-\det h_{ab}} = {R\over\gamma_\perp(\sigma)} \left(1 -
2\sigma^2 R^2 \gamma_\perp^{2}(\sigma)\omega\dot\phi\right)\, .
\end{eqnarray}
In the above expressions we have used the notation
\begin{eqnarray}
\gamma_\perp^{-2}(\sigma) &=& 1 - \sigma^2\omega^2 R^2\, , \\
\gamma^{-2}(\sigma) &=& \gamma_\perp^{-2}(\sigma) - \sigma^2 \dot{R}^2\,
. \label{gammaFactors}
\end{eqnarray}

\begin{figure}[htbp]
\hbox{\epsfxsize = 5 in \hskip 1.20 in \epsfbox{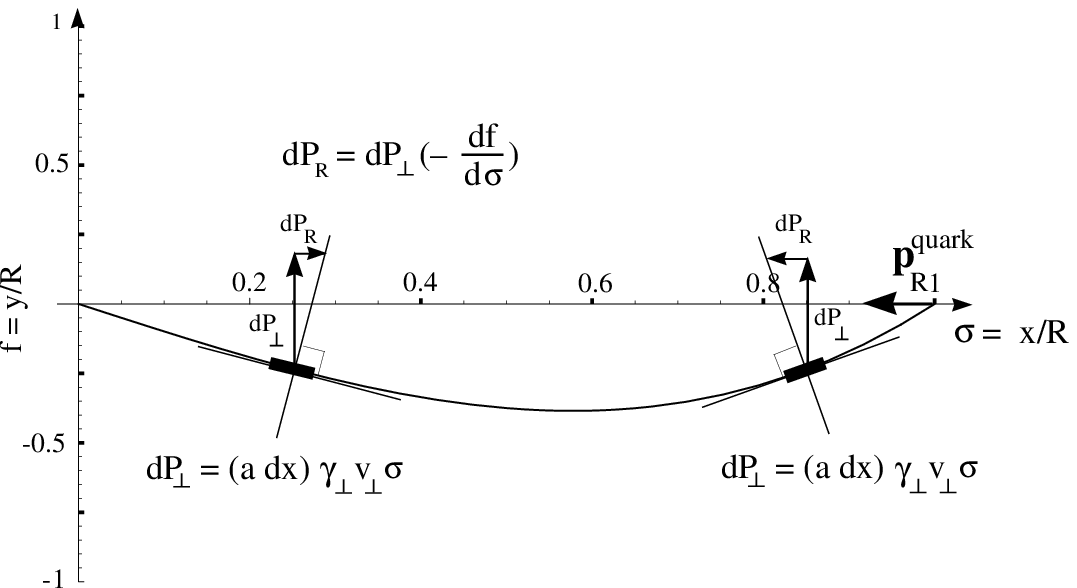}}\vskip 1 cm
\caption{The solution to the string shape (\protect\ref{Xperpshape}) with
positive $\dot\omega$ and $v_\perp \ll 1$.  The transverse and radial
momenta of two elements are depicted. The contribution at the larger radius
will predominate giving a net radial momentum parallel to the quark's
radial momentum.}
\label{fig:one}
\end{figure}

The equations of motion for the string (\ref{StringEoM}) hold both for the
temporal $X^0 = t$ and spatial $X$ components.  In the case of $X^0$, we
have, to first order in $f$ and $\phi$
\begin{equation}
\left[ \gamma_\perp(\sigma) + \gamma_\perp^3(\sigma) \sigma^3 \omega
\dot\phi\right]_{,t} = \left[\gamma_\perp(\sigma)\sigma\dot{R} +
\gamma_\perp(\sigma) \sigma^2{R} \omega f^\prime\right]_{,\sigma} \,
. \label{pertEoM}
\end{equation}
Using the easily verified identity $(\gamma_\perp(\sigma) R)_{,t} =
 \dot{R}(\sigma \gamma_\perp(\sigma))_{,\sigma}\,$, and dropping
 $\dot{R}\dot\phi$ terms, we reduce Eq.~(\ref{pertEoM}) to
\begin{equation}
{d\over d\sigma}\left(\sigma^2\gamma_\perp(\sigma) {d f\over
d\sigma}\right) = R^2 \dot\omega\sigma^2\gamma_\perp^3(\sigma)\,
.\label{df}
\end{equation}
Here we introduce the notation $\dot\omega\equiv\ddot\phi$.  It can be
shown \cite{adiabatic} that the equations of motion for the spatial
coordinate $X^i$ also reduce to this same equation.  Integration of
Eq.~(\ref{df}) yields
\begin{eqnarray}
{df\over d\sigma} &=& {R^2\dot\omega\over v^2_\perp}\left[\,\sigma -\,
{\arcsin \sigma v_\perp\over \sigma^2 \gamma_\perp(\sigma)}\right] + C_1
{1\over \sigma^2\gamma_\perp(\sigma)}\,\, ,
\end{eqnarray}
whose solution is 
\begin{eqnarray}
\sigma f(\sigma) &=& {\dot\omega R^2\over v^3_\perp} \left[{1\over2}\sigma
v_\perp \arcsin \sigma v_\perp\right]\arcsin\sigma v_\perp \nonumber \\ &&
+ \,\, C_1 \sigma v_\perp + C_2 \left(\sqrt{1-\sigma^2v_\perp^2} + \sigma
v_\perp\arcsin\sigma v_\perp\right)\, .
\end{eqnarray}
In the above we have defined the transverse end velocity $v_\perp \equiv
\omega R$.  After removal of the rigidly rotating phase, the string
coordinate can be written in terms of a radial and a transverse piece
\begin{equation}
{\bf X} = {1\over\sqrt2}\left(X_R +  i X_\perp\right)\, , 
\end{equation}
which are given by
\begin{eqnarray}
{X_R\over R} &=& \sigma \, , \\
{X_\perp\over R} &=& \sigma f(\sigma) \equiv {1\over 6}\dot\omega R^2\, {\rm
shape}(\sigma, v_\perp)\, .\label{Xperpshape}
\end{eqnarray}
Imposing the end conditions $X_\perp(\sigma = 0) = X_\perp(\sigma=1) = 0$,
we find
\begin{eqnarray}
{\rm shape}(\sigma, v_\perp) &=& -\, {6\over v_\perp^3}\Bigg[ {1\over2}\sigma
v_\perp \left( (\arcsin v_\perp)^2 - (\arcsin \sigma v_\perp )^2 \right)
\nonumber \\ && +\, \sigma\,\sqrt{1-v_\perp^2}\arcsin v_\perp -
\sqrt{1-\sigma^2 v_\perp^2} \arcsin \sigma v_\perp\Bigg]\, . \label{shapefcn}
\end{eqnarray}
The string shape for $v_\perp \ll 1$ is shown in Fig.~\ref{fig:one}.  Other
examples are shown in Ref.~\cite{adiabatic}.

As we pointed out earlier, one can compare the angular momentum, energy and
linear momentum of the curved and straight string.  The results are
\begin{eqnarray}
J^3_{\rm curved} &=& J^3_{\rm straight} = {a R^2\over 2
v_\perp}\left[{\arcsin(v_\perp)\over v_\perp} - \sqrt{1 -
v_\perp^2}\right]\, , \label{Jcurved} \\ 
E_{\rm curved} &=& E_{\rm straight} = a R\,
{\arcsin{v_\perp}\over v_\perp}\ , \label{Ecurved} \\ 
P_{\perp\, \rm curved} &=& P_{\perp\,
\rm straight} = {a R \over v_\perp}\left[1 - \sqrt{1-v_\perp^2}\right]\ .
\end{eqnarray}
Although there is no radial tube momentum for the straight string, it is
non-zero for the curved string.  This is evident from Fig.~1 and by direct
calculation 
\begin{equation}
P_R = -\,a R { \dot\omega R^2 \over v_\perp^4}\left[1 - \sqrt{1-v_\perp^2}
- {v_\perp\over 2}\arcsin v_\perp\right] \arcsin v_\perp\ .\label{PsubR}
\end{equation}
The presence of radial string momentum is the only dynamical evidence of
the curved string.

We have considered so far in this section only the heavy-light meson case
where one tube end is fixed.  Although the general straight string case was
discussed previously \cite{aft} we will consider here the generalization to
curved strings of the equal quark mass case.  In Fig.~2 we show the shape
function from $-1\le\sigma\le +1$ which describes the equal mass case.  It
is evident that we may immediately find the angular momentum, energy, and
linear momentum from the preceding expressions by using the substitution
\begin{equation}
\begin{array}{rcl}
R &\rightarrow& r/2\,\, ,  \\
v_\perp = \omega R &\rightarrow& v_\perp = \omega r/2\,\, ,
\end{array}
\end{equation}
which leads to the following expressions for the equal quark mass case
\begin{eqnarray}
J^3_{\rm equal\ mass} &=& {a r^2\over 4 v_\perp}\left[{\arcsin v_\perp \over
v_\perp} - \sqrt{1 - v_\perp^2}\right]\, , \label{Jem} \\ 
E_{\rm equal\ mass} &=& a r\, {\arcsin{v_\perp}\over v_\perp}\ ,
\label{Eem} \\ 
P_r &=& -\,{ ar}\,\, {\dot\omega r^2 \over 8 v_\perp^4}\left[1 -
\sqrt{1-v_\perp^2} - {v_\perp\over 2}\arcsin v_\perp\right] \arcsin
v_\perp\ \label{Prem}.
\end{eqnarray}
Again we note that $E$ and $J^3$ for the straight and curved strings are
identical.  In the above we increased $E$ and $J^3$ by a factor of two
because of the two string pieces.  Of course we must add the quark
contribution to each quantity to obtain the total.

\section{Quantization and Method of Numerical Solution}\label{sec:numerical}

We begin by reviewing the quantization of the classical expressions for the
heavy-light straight string. From (\ref{spatialP}), (\ref{Jcurved}) and
(\ref{stringE}), (\ref{Ecurved}) the angular momentum and energy are
\begin{eqnarray}
J^3 &\equiv& J = m \gamma v_\perp R + a R^2 F(v_\perp)\, , \label{Jthree} \\
E &=& M_Q + m\gamma + a R \, {\arcsin v_\perp \over v_\perp}\, ,\label{Esys}
\end{eqnarray}
where $M_Q$ is the heavy quark mass and $F(v_\perp)$ is defined as
\begin{eqnarray}
F(v_\perp) &=& {1\over 2v_\perp}\left({\arcsin v_\perp\over v_\perp} -
\sqrt{1 - v_\perp^2}\right)\, .
\end{eqnarray}
In the above $v_\perp = \omega R$ and we have used the notation
\begin{eqnarray}
\gamma_\perp &=& {1\over \sqrt{1 - v_\perp^2}}\,\, , \\
\gamma &=& {1\over \sqrt{1 - v_\perp^2 - \dot{R}^2}}\,\, .
\end{eqnarray}
The radial quark velocity $\dot R$ has to be eliminated in favor
of the radial quark momentum by the relation
\begin{eqnarray}
({p}^{\rm quark}_R)^2 + m^2  = m^2 \left({\dot R^2\over 1 - v_\perp^2 -
\dot{R}^2} + 1\right) = {(m\gamma)^2\over \gamma_\perp^2}\, , 
\end{eqnarray}
which leads to 
\begin{eqnarray}
m \gamma &=& W_R \gamma_\perp\, , \label{mgamma}\\
W_R &=& \sqrt{({p}^{\rm quark}_R)^2 + m^2 }\, \label{wr-straight}.
\end{eqnarray}
From (\ref{Jthree}), (\ref{Esys}), and (\ref{mgamma}) we have
\begin{eqnarray}
{J\over R} &=& W_R \gamma_\perp v_\perp + aR F(v_\perp)\, , \label{Jemc} \\
E &=& M_Q + W_R \gamma_\perp + aR {\arcsin v_\perp \over v_\perp}\,
\label{Eemc} .
\end{eqnarray}

The above equations can be quantized \cite{aft} by making the replacements
\begin{equation}
J \rightarrow \sqrt{\ell(\ell+1)}, \ \ \ \
p_R^2\rightarrow -\frac{1}{R}\frac{\partial^2}{\partial R^2}R\ ,
\end{equation}
and by symmetrizing terms which depend on both $R$ and $v_\perp$ with
anticommutators ($\{A,B\}\equiv AB+BA$). This
procedure leads to the following quantum-mechanical
equation for the angular momentum and the energy 
\begin{eqnarray}
\frac{\sqrt{\ell(\ell+1)}}{R}&=&\frac{1}{2}\{W_R,\gamma_\perp v_\perp \} 
+ a \{R,F(v_\perp)\}\ ,
\label{hlJ}
\\
H &=& M_Q + \frac{1}{2}\{W_R,\gamma_\perp\}+
\frac{a}{2}\{R,\frac{\arcsin{v_\perp}}{v_\perp}\}\, . 
\label{hlH}
\end{eqnarray}

As we have shown in \cite{adiabatic}, the only difference between the
straight and the curved QCD string is that the later has radial momentum,
\begin{equation}
p_R^{\rm tube} = -\,a R ( \dot\omega R^2 ) g(v_\perp)\ , \label{Ptc}
\end{equation}
where we defined
\begin{equation}\label{gvperp}
g(v_\perp)= \left(1 - \sqrt{1-v_\perp ^2} - {v_\perp \over 2}\arcsin
v_\perp \right) {\arcsin v_\perp \over v_\perp^4} \ .
\end{equation}
Since $p_R^{\rm quark} = p_R - p_R^{\rm tube}$, Eq.~(\ref{wr-straight})
then becomes,
\begin{equation}
\label{wr-curved}
W_R=\sqrt{(p_R-p_R^{\rm tube})^2+m^2}\ .
\end{equation}

In order to quantize the above expressions it is convenient to write
\begin{eqnarray}
\dot \omega R^2 &=& \dot{v}_\perp R -  \dot R v_\perp \nonumber \\
&=& \frac{1}{2}\left(\{\dot{v}_\perp, R\} -  \{\dot R,v_\perp\}\right) \ .
\label{dot-omega-r2}
\end{eqnarray}
After promoting $v_\perp$ and $R$ to quantum-mechanical operators in
(\ref{dot-omega-r2}) we can use the quantum dynamical equations of motion
\begin{eqnarray}
\dot v_\perp &=& -i [v_\perp,H]\ ,\\
\label{Rdot} \dot R &=& -i [R,H]\ ,
\end{eqnarray}
which completely specifies the quantum version of $\dot\omega R^2$ in terms
of $R$, $v_\perp$, and Hamiltonian given in (\ref{hlH}), and allows us to
quantize $p_R^{\rm tube}$ by choosing appropriate symmetrization between
the non-commuting operators. This procedure gives us
\begin{equation}
p_R^{\rm tube} = -\,a\, 
\frac{1}{2}\left\{ 
\dot\omega R^2,
\frac{1}{2}\{R, g(v_\perp)\}
\right\} \ .
\label{prt}
\end{equation}
The final step in quantizing (\ref{wr-curved}) is to expand
$(p_R-p_R^{\rm tube})$ by writing
\begin{equation}
(p_R-p_R^{\rm tube})^2 = p_R^2 - \{p_R, p_R^{\rm tube}\} + ({p_R^{\rm
tube}})^2\ .\label{PRquark}
\end{equation}
Here, $p_R=-\frac{i}{R}\frac{\partial}{\partial R} R$ and $p_R^{\rm tube}$
is defined by (\ref{dot-omega-r2}) through (\ref{prt}).

The solution method for the straight string equations is discussed in
detail in Ref.~\cite{aft}. The fundamental problem is that the unknown
operator $v_\perp$ cannot be analytically eliminated, so one must eliminate
it numerically. This can be accomplished by expanding the radial wave
function in terms of a complete set of basis states, which reduces
(\ref{hlJ}) to a transcendental infinite dimensional matrix equation.  By
truncating the number of the basis states to a finite number $N$, the
matrix equation becomes finite dimensional, and can be solved by iteration
procedure described in \cite{aft}. Once the matrix for $v_\perp$ is known,
one can calculate the Hamiltonian matrix from (\ref{hlH}) and solve for the
energy eigenvalues and corresponding wave functions.  The modified
algorithm for solving the curved string equations involves continually
updating the $W_R$ matrix during the iteration procedure \cite{aft} in
which the matrix $v_\perp$ is found.
  
Note that truncation introduces the dependence of the eigenvalues on the
variational parameter $\beta$ describing the basis states. However, if a
sufficiently large number of basis states are used, this dependence will
exhibit a ``plateau'' indicating a stable solution.  Examples of this will
be shown in Sec.\ \ref{sec:results}.

The equal quark mass meson can be quantized by a simple modification of the
above.  In this case we start with Eqs.~(\ref{Jem})-(\ref{Prem}) to obtain
expressions analogous to (\ref{hlJ}), (\ref{hlH}), and (\ref{prt})
\begin{eqnarray}
\label{J-v-perp} {\sqrt{\ell(\ell+1)}\over r} &=& \Big\{W_r,\gamma_\perp
v_\perp\Big\} + \frac{a}{4} \Big\{r, F(v_\perp)\Big\}\,\, , \\
\label{E-v-perp} H &=& \Big\{ W_r, \gamma_\perp\Big\} + {a\over2} \Big\{r,
{\arcsin v_\perp\over v_\perp}\Big\}\,\, , \\ \label{Pr-v-perp} p^{\rm
tube}_r &=& -\, {a\over 16} \Big\{ \dot\omega r^2, {1 \over 2} \Big\{ r,
g(v_\perp) \Big\} \Big\}\,\, .
\end{eqnarray}
The numerical solution is step by step the same as for the heavy-light
case.

The more general case of unequal arbitrary masses at the ends is one step
more involved.  This case was solved in \cite{aft} for the straight string.
For mesons with unequal mass quarks, one must require that $P_\perp^{\rm
tot} \equiv 0$.  This is equivalent to solving for the center of momentum
in addition to the angular momentum and energy equations.  The curved
string dynamical solution should proceed similarly to the two special cases
discussed previously.

\section{Numerical Results}\label{sec:results}

For zero angular momentum states we observe from (\ref{hlJ}) and
(\ref{J-v-perp}) that $v_\perp = 0$ as expected.  The tube radial momentum
is zero and the Hamiltonians (\ref{hlH}) and (\ref{E-v-perp}) become
equivalent to ``spinless Salpeter'' equations, which have well defined
numerical solutions \cite{aft}.  We will therefore confine ourselves to
non-zero angular momenta.

The results for equal mass and heavy-light mesons are qualitatively similar
and we will discuss these string configurations in parallel, pointing out
the similarities and differences.  As mentioned in the preceding section
the set of basis states have a scale parameter $\beta$.  For a given
truncation to $N$ basis states it is customary to plot energy eigenvalues
as a function of $\beta$ to identify a ``plateau'' indicating a stable
solution.  In Fig.~\ref{fig:three} (heavy-light) and Fig.~\ref{fig:four}
(equal mass) we show the $\beta$ dependence of the excitation energy of the
$p$-wave ground state meson with massless light quarks.  The energy
calculation is done twice; first with the straight string model and then
with the curved string.  We note that as the number of basis states
increases definite plateaus develop.

The second observation we make from Figs.~\ref{fig:three} and
\ref{fig:four} is that the curved string energy is slightly less than the
straight string energy.  We believe this results from the reduction of the
effective string tension due to the transverse motion of the string during
the quark's motion.  The difference in the $p$-wave case, shown in
Fig.~\ref{fig:three}, is just a few $MeV$ but this varies with angular
momentum and radial excitation.

In Figs.~\ref{fig:five} and \ref{fig:six} we illustrate how the curved and
straight string differ for a variety of low lying states.  The cases shown
are for zero mass light quark(s) at the end(s) of the string.  The general
trend is that the difference increases from zero at $\ell = 0$ (which must
be the case) to about $15~MeV$ at the higher angular momentum values shown.

The dependence of the energy difference for $p$-wave solutions on the mass
of the quark at either end of the string is illustrated in
Fig.~\ref{fig:seven} (heavy-light) and Fig.~\ref{fig:eight} (equal mass).
The energy difference $\Delta E$ is seen to decrease rapidly as the quark
mass increases.  As expected, the curvature effect on the solution is
largest for massless quarks.

Finally, in Fig.~\ref{fig:nine} we show the Regge pattern for the curved
string with one massless quark and the other end fixed. We exhibit the
corresponding Regge pattern for massless quark ends in Fig.~\ref{fig:ten}.
On this scale they are similar to those of the straight string.  One
observes towers of nearly degenerate states of even (or odd) angular
momentum.  The two cases are analogous except that the heavy-light Regge
slope is double the light-light slope. Except for the value of the slope,
this pattern is very similar to that of scalar potential confinement
\cite{mgo,observations}.

\section{Conclusion}\label{sec:conclusion}

In this paper we have investigated the dynamical effects of the small
radial momentum developed by a QCD string as it bends in response to the
angular acceleration of the quarks on the ends.  Our principal results are
that one can successfully implement the curved string solutions into a
dynamical meson model with massive quarks at the string ends.  The solution
is self-consistent in that for a physical state the deviation from
straightness is small.  For such small transverse string motion the energy,
angular momentum, and transverse momentum are identical to those of a
straight string with the same angular velocity.  When the small curvature induced radial momentum is
incorporated into the numerical method one finds solutions that are very
similar to those of the straight string.

The curvature effects are largest for massless quarks. Even in this case
one finds shifts of at most a few $MeV$.  Our results indicate that for
most purposes the simpler straight string model gives nearly correct
results and if desired, more accuracy can be readily achieved.

\section*{Acknowledgments}
This work was supported in part by the US Department of Energy under
Contract No.~DE-FG02-95ER40896.  Fermilab is operated by URA under DOE
contract DE-AC02-76CH03000.

\begin{figure}[htbp]
\epsfxsize=\hsize
\hbox{\hskip 0 in \epsfbox{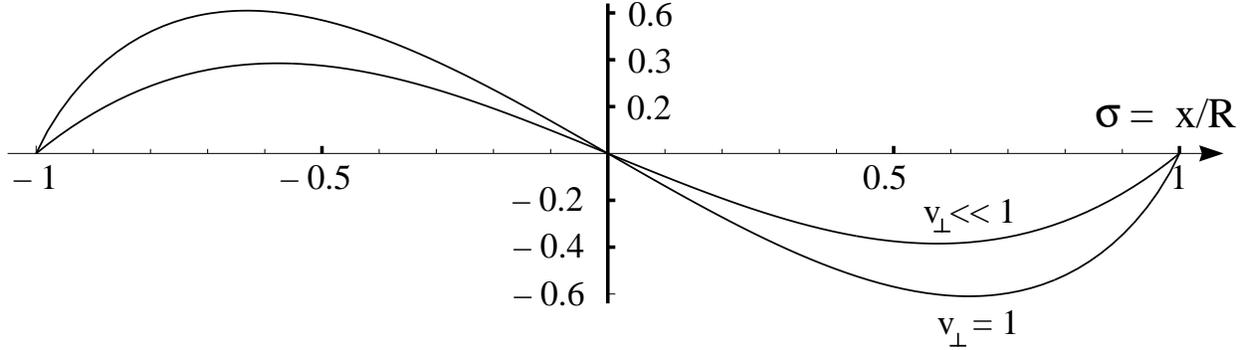}}\vskip 1 cm
\caption{The solution to the string shape (\protect\ref{Xperpshape}) with
positive $\dot\omega$ for a string with equal mass quarks at its ends.  The
``shape function'' (\protect\ref{shapefcn}) is shown for non-relativistic
and ultra-relativistic quark rotation. }
\label{fig:two}
\end{figure}

\begin{figure}
\epsfxsize = \hsize 
\epsfbox{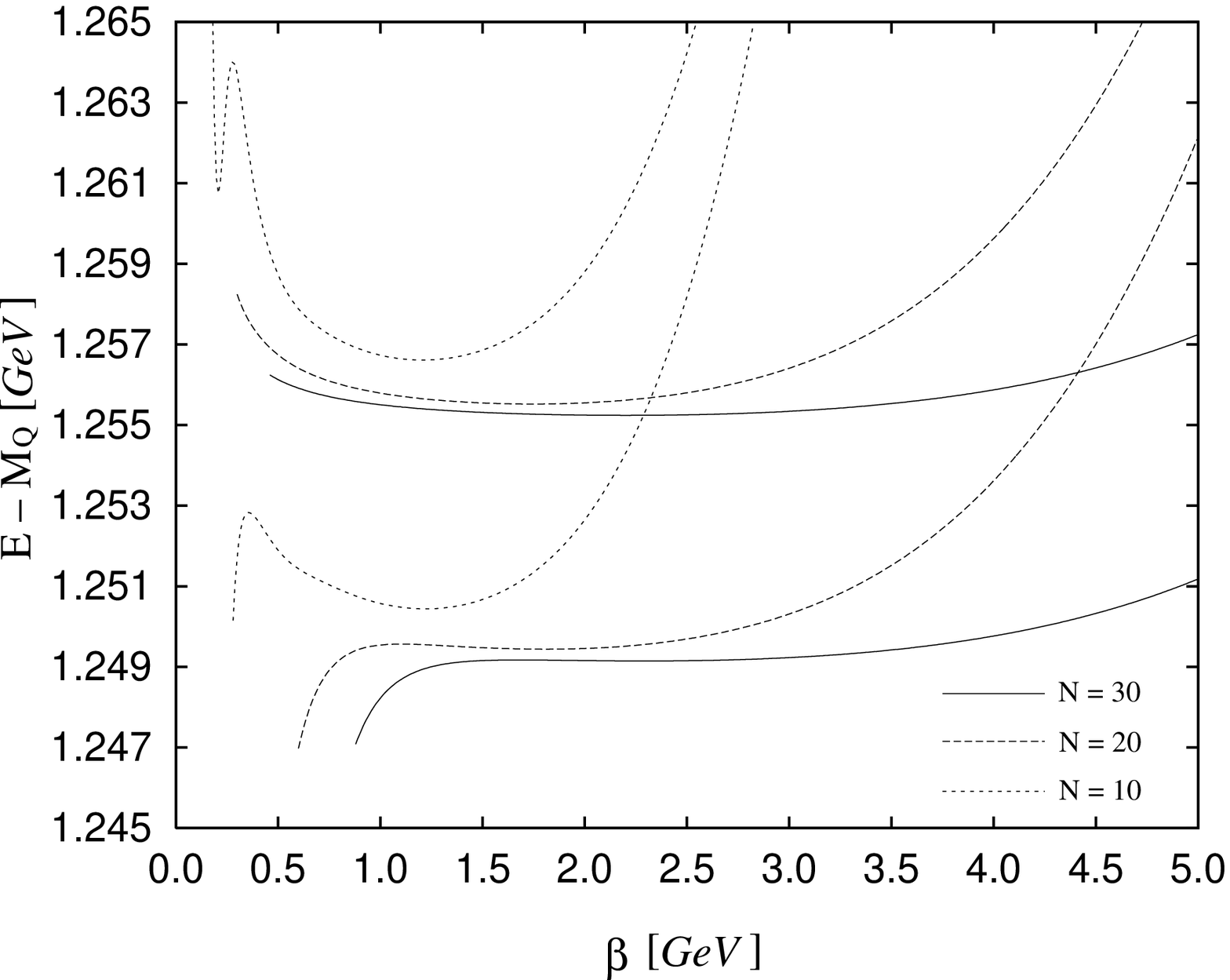}\vskip 1 cm
\begin{caption}
{Dependence on the variational parameter $\beta$ of the ground state
$p$-wave excitation energy of the heavy-light Hamiltonian with the straight
string (upper curves) and with the curved string (lower curves). We used $a
= 0.2~GeV^2$ and light quark mass $m=0$.  Results shown were obtained with
10, 20 and 30 basis states.}\label{fig:three}
\end{caption}\end{figure}

\begin{figure}
\epsfxsize=\hsize
\epsfbox{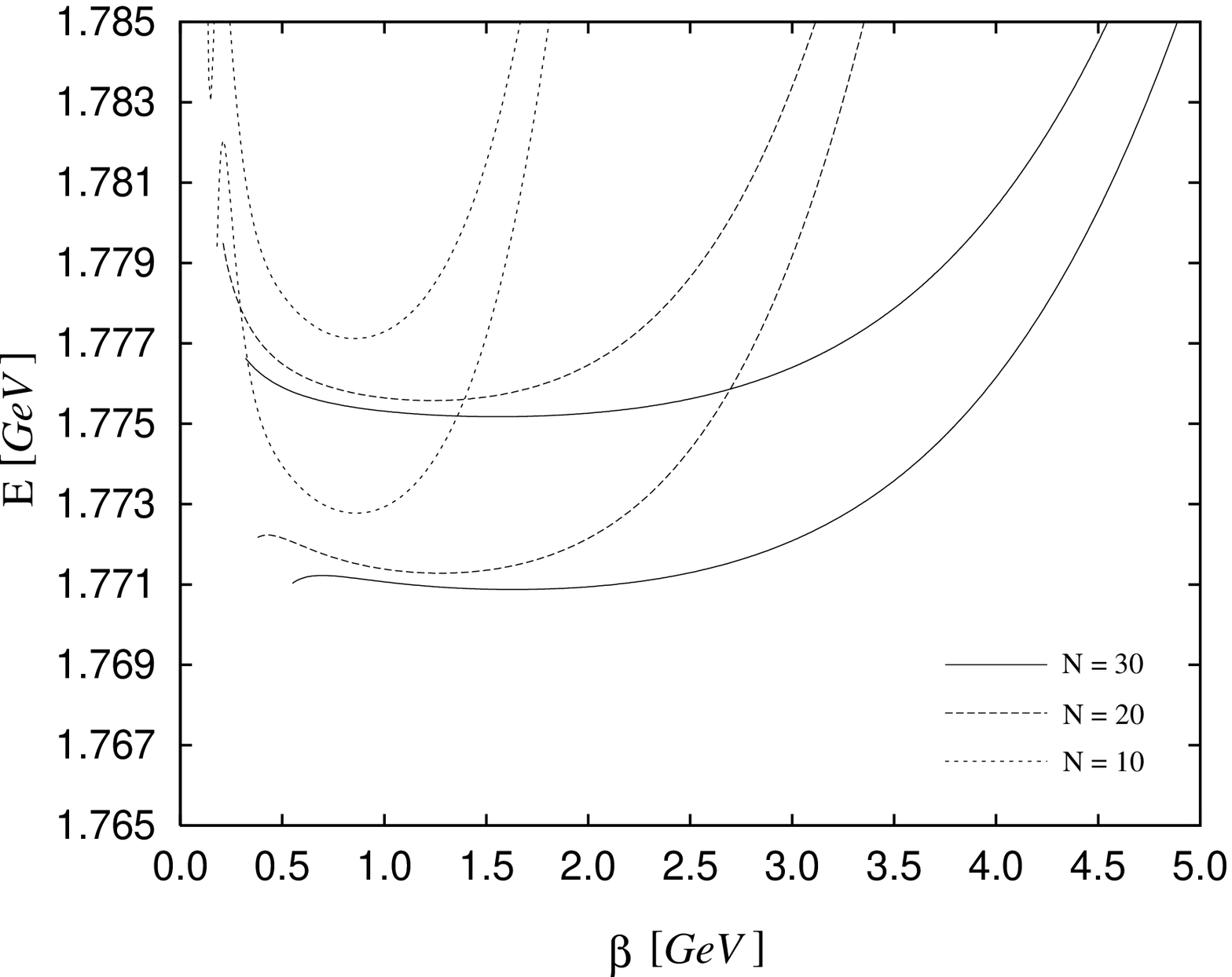}\vskip 1 cm
\begin{caption}
{Dependence on the variational parameter $\beta$ of the ground state
$p$-wave excitation energy of the equal mass Hamiltonian with the straight
string (upper curves) and with the curved string (lower curves). We used $a
= 0.2~GeV^2$ and quark mass $m=0$.  Results shown were obtained with 10, 20
and 30 basis states.}\label{fig:four}
\end{caption}\end{figure}

\begin{figure}
\epsfxsize=\hsize
\epsfbox{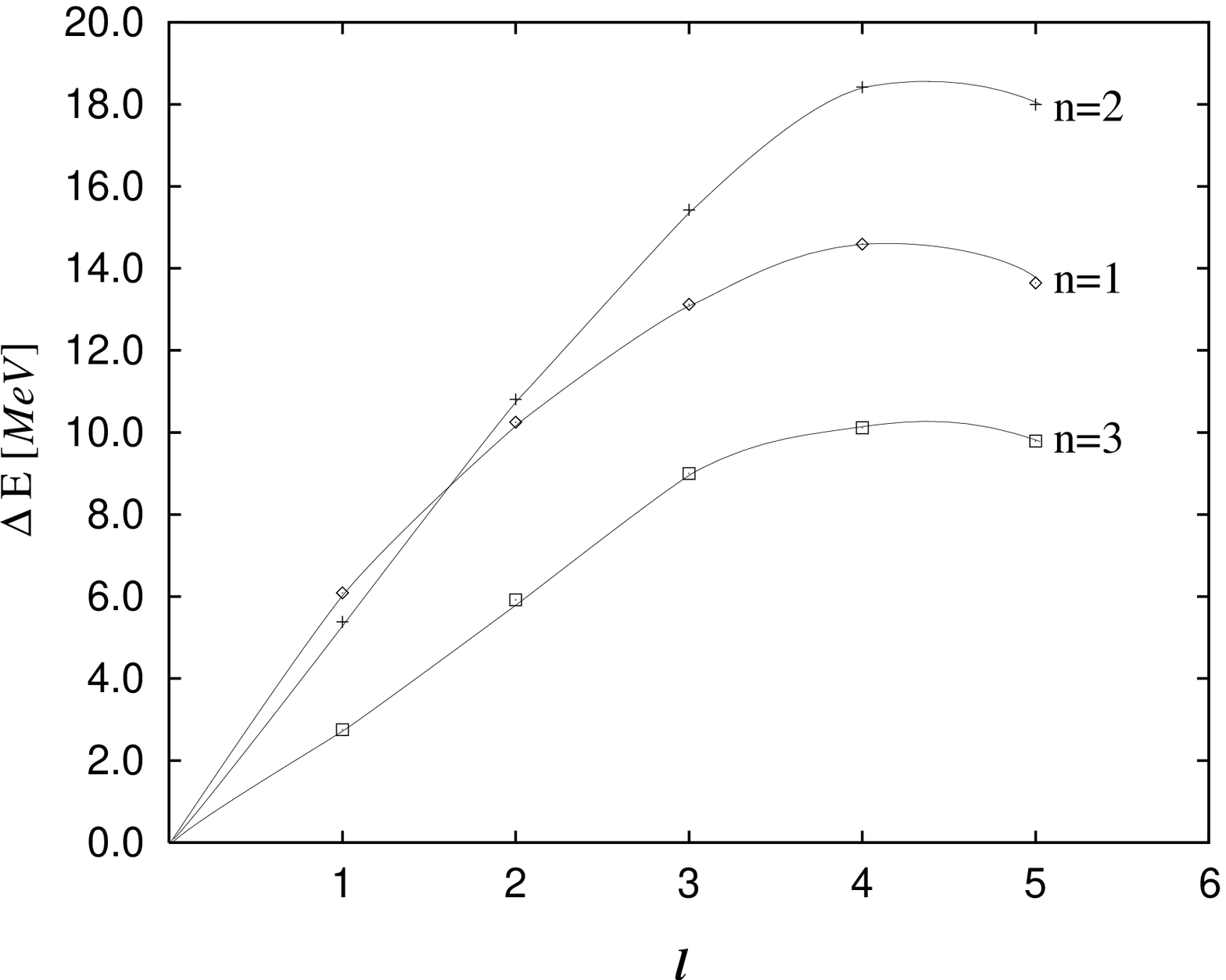}\vskip 1 cm
\begin{caption}
{Difference between the energies of the straight and curved heavy-light
Hamiltonian as a function of the orbital angular momentum.  We used $a =
0.2~GeV^2$ and light quark mass $m=0$.  Results shown are for the ground
state ($n=1$) and for the first two radially excited
states. }\label{fig:five}
\end{caption}\end{figure}

\begin{figure}
\epsfxsize=\hsize
\epsfbox{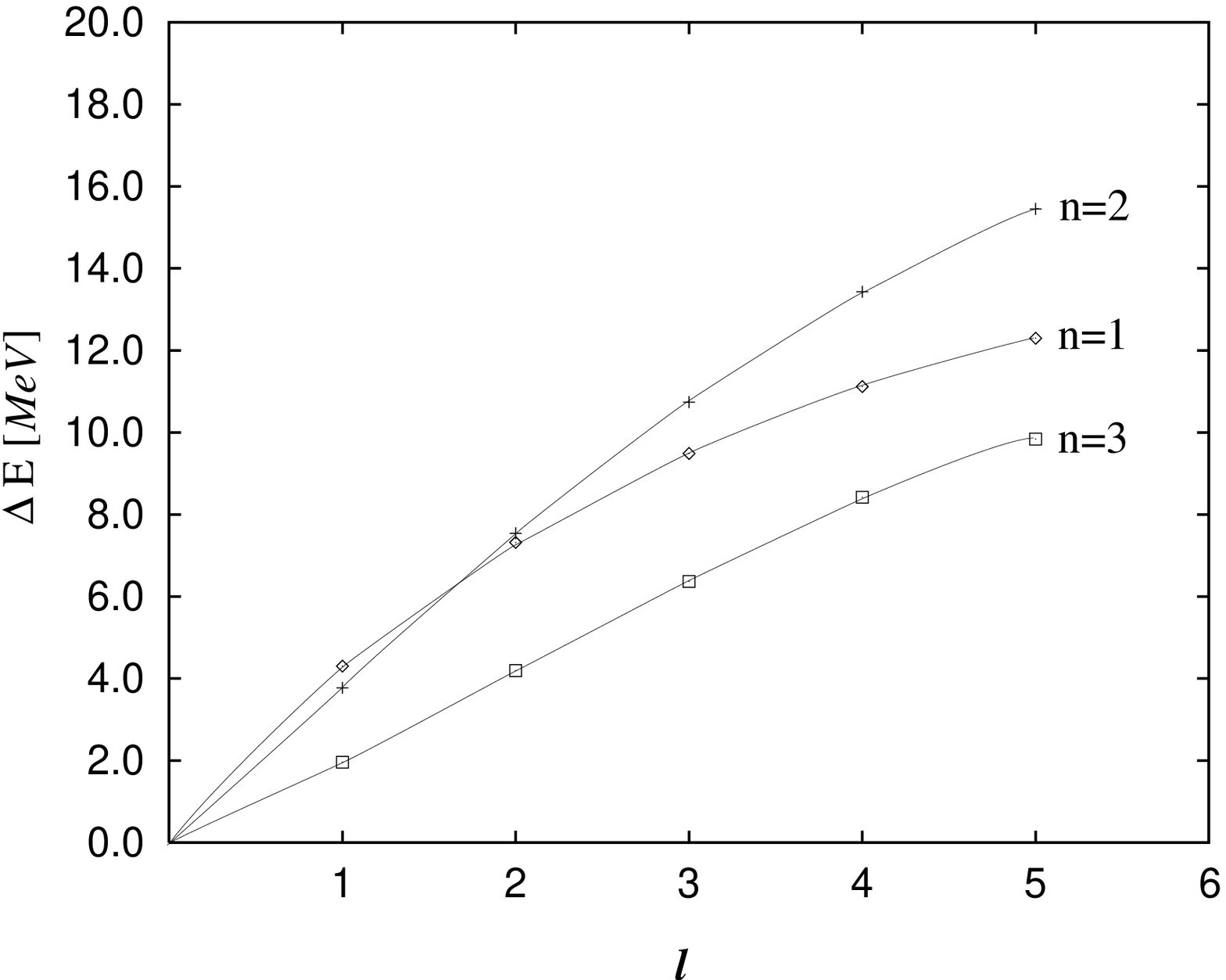}\vskip 1 cm
\begin{caption}
{Difference between the energies of the straight and curved equal mass
Hamiltonian as a function of the orbital angular momentum.  We used $a =
0.2~GeV^2$ and light quark mass $m=0$.  Results shown are for the ground
state ($n=1$) and for the first two radially excited
states. }\label{fig:six}
\end{caption}\end{figure}

\begin{figure}
\epsfxsize=\hsize
\epsfbox{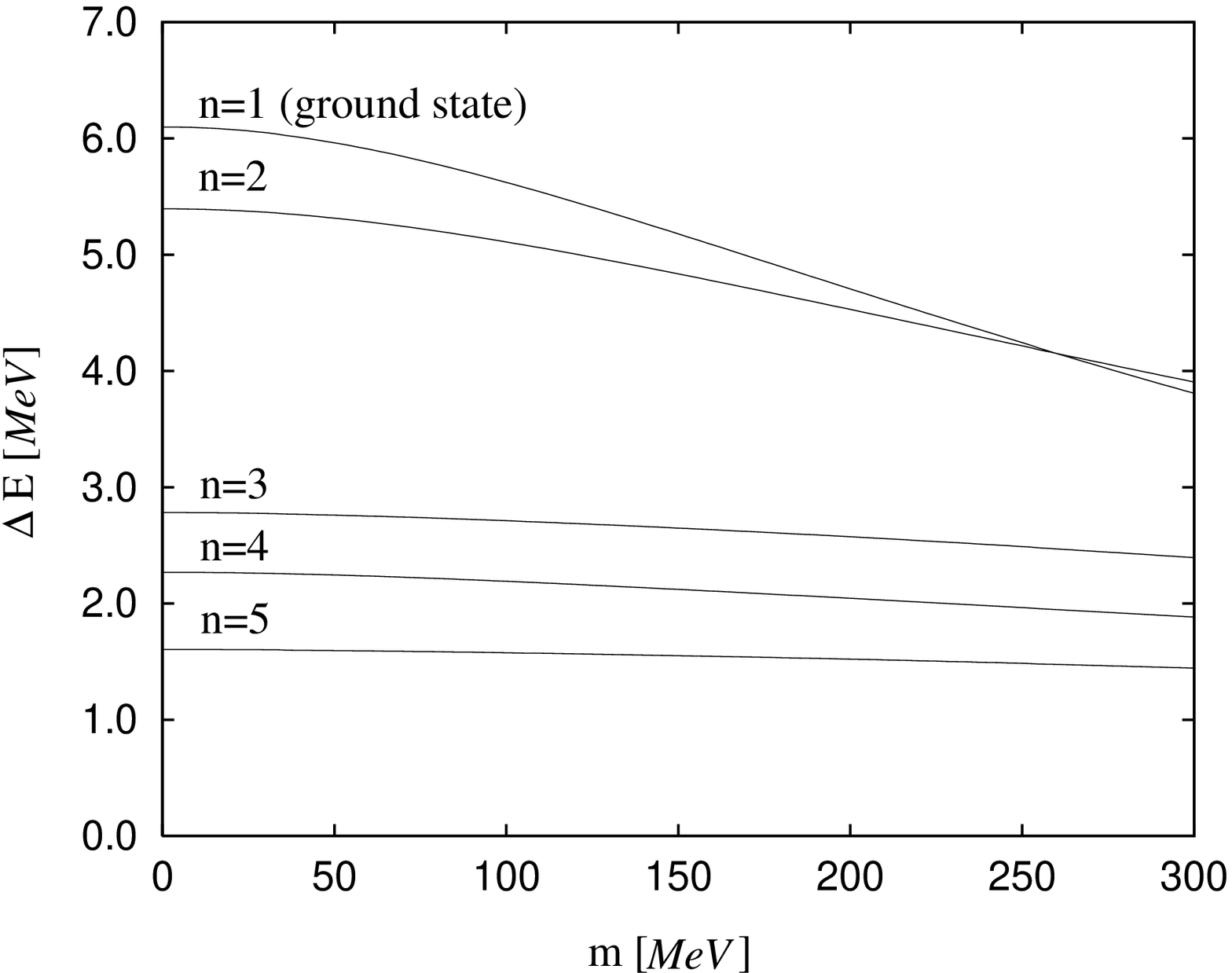}\vskip 1 cm
\begin{caption}
{Difference between the $p$-wave ($\ell=1$) energies of the straight and
curved heavy-light Hamiltonian as a function of the light quark mass.  We
used $a = 0.2~GeV^2$, and results shown are for the ground state ($n=1$)
and for the first four radially excited states.}\label{fig:seven}
\end{caption}\end{figure}

\begin{figure}
\epsfxsize=\hsize
\epsfbox{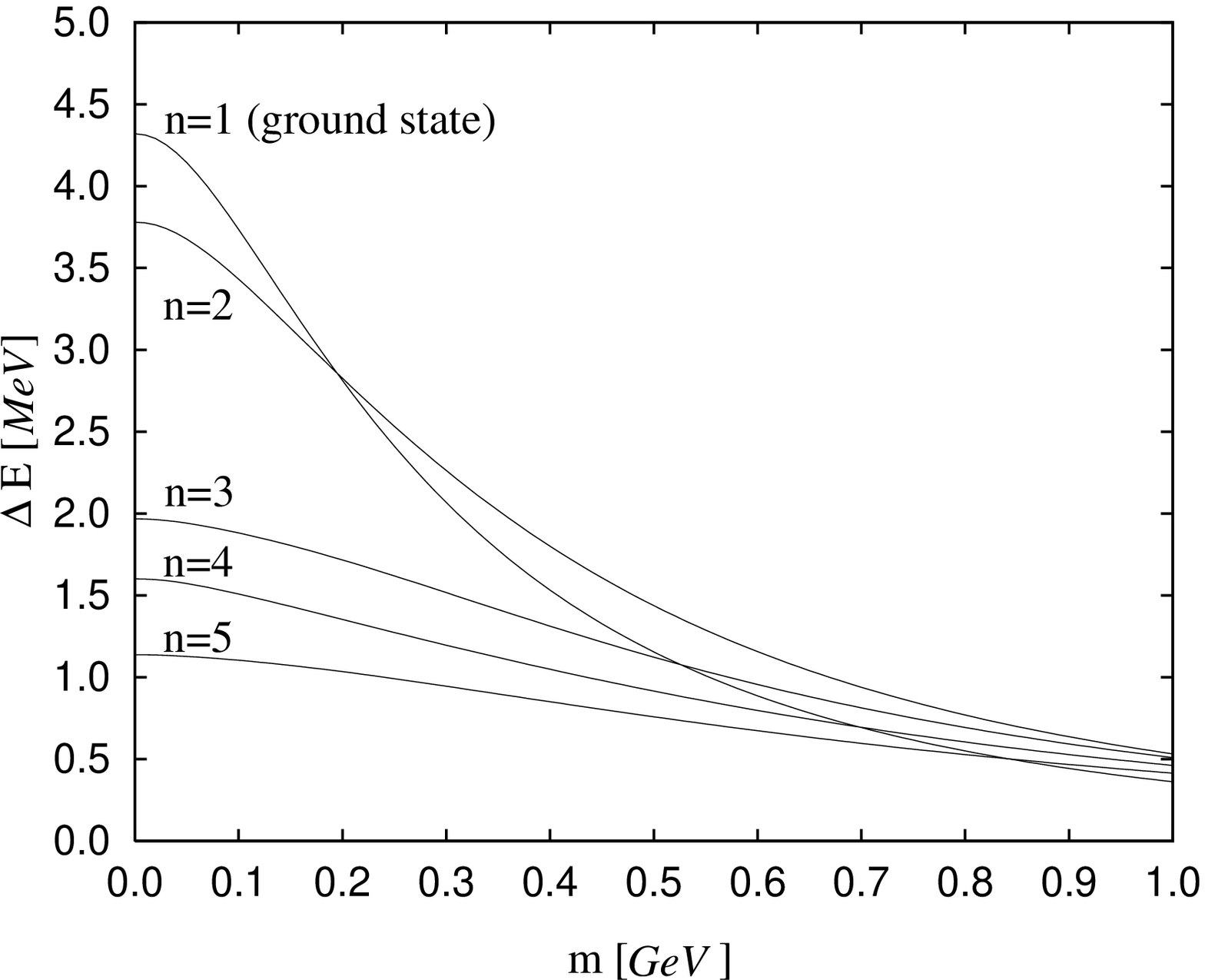}\vskip 1 cm
\begin{caption}
{Difference between the $p$-wave ($\ell=1$) energies of the straight and
curved equal-mass Hamiltonian as a function of the quark mass.  We used $a
= 0.2~GeV^2$, and results shown are for the ground state ($n=1$) and for
the first four radially excited states.}\label{fig:eight}
\end{caption}\end{figure}

\begin{figure}
\epsfxsize=\hsize
\epsfbox{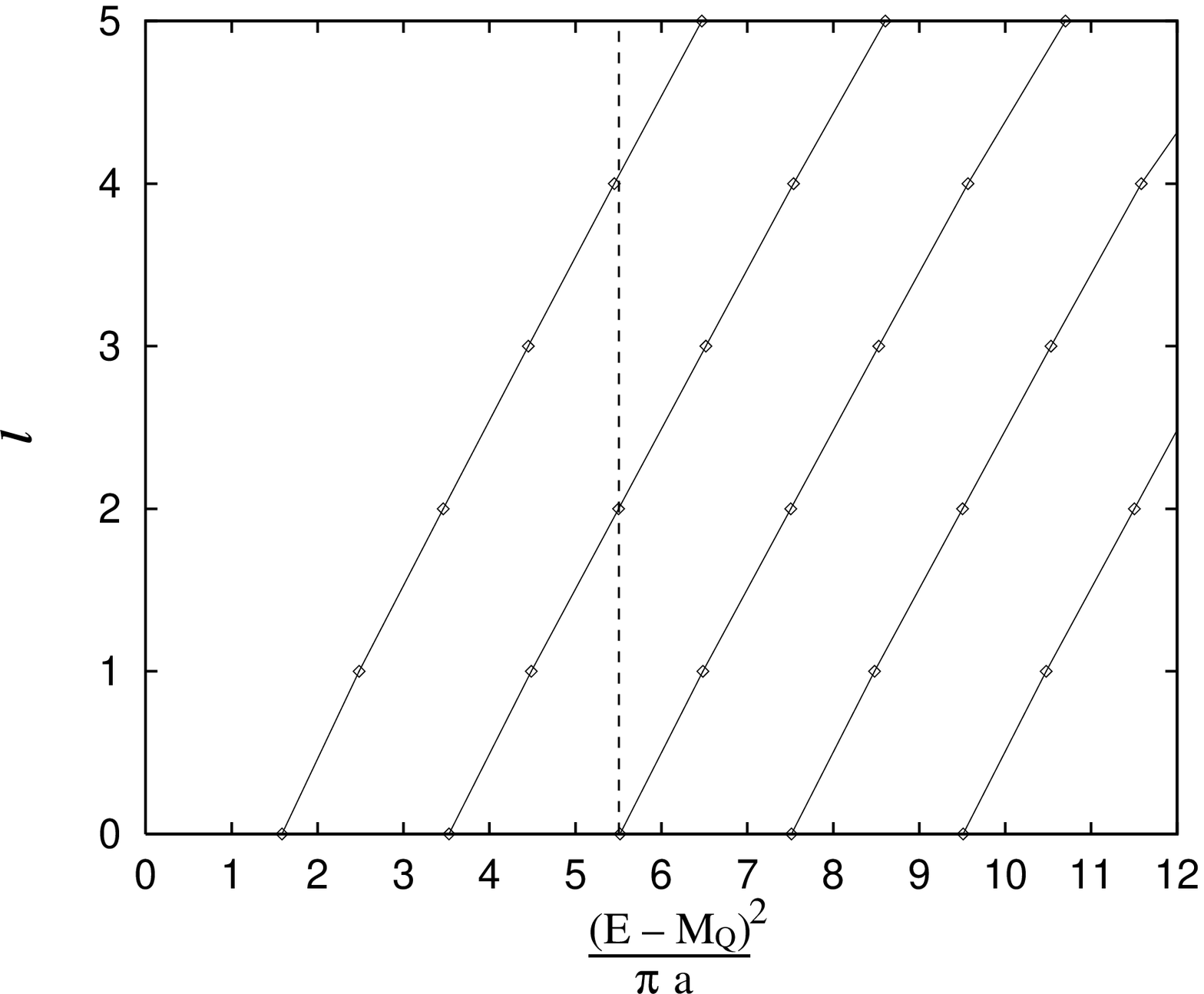}\vskip 1 cm
\begin{caption}
{Regge trajectories for the heavy-light Hamiltonian with the curved string
having a massless quark.  The vertical line illustrates the tower structure
of nearly degenerate states of alternating angular
momentum.}\label{fig:nine}
\end{caption}\end{figure}

\begin{figure}
\epsfxsize=\hsize
\epsfbox{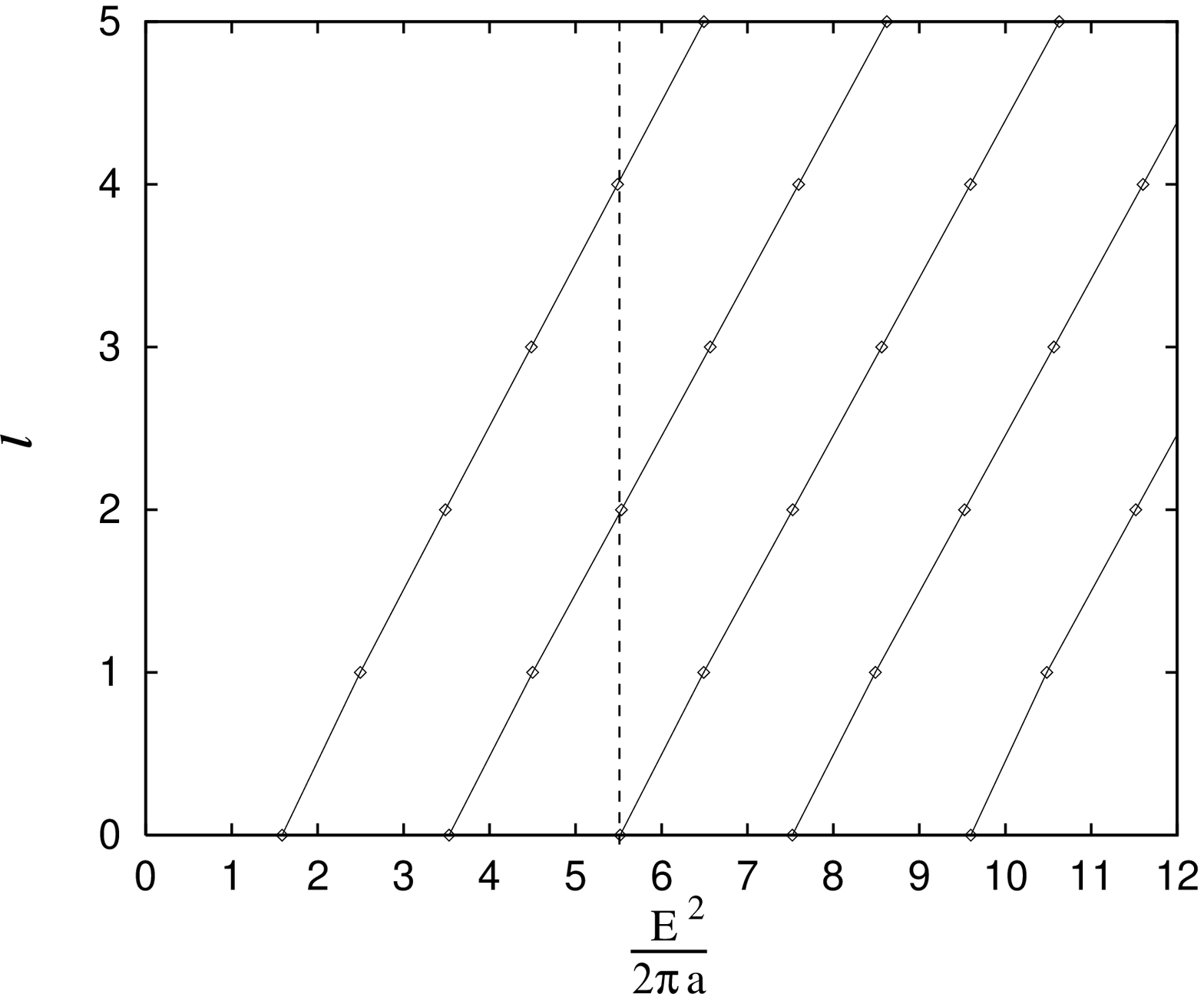}\vskip 1 cm
\begin{caption}
{Regge trajectories for tphe equal-mass Hamiltonian with the curved string
having massless quarks at its ends.  The vertical line illustrates the
tower structure of nearly degenerate states of alternating angular
momentum.}\label{fig:ten}
\end{caption}\end{figure}

\end{document}